\documentclass[letterpaper,11pt]{article}
\usepackage[margin=1in]{geometry}
\usepackage{amsfonts,amsmath,amssymb,amsthm,thmtools,mathtools}
\usepackage{booktabs}
\usepackage{enumitem}
\usepackage[dvipsnames]{xcolor}
\usepackage{hyperref}
\usepackage{cleveref}

\hypersetup{
    colorlinks=true,
    linkcolor=RoyalBlue,
    citecolor=ForestGreen,
    urlcolor=RoyalBlue,
    hypertexnames=false
}

\declaretheorem[numberwithin=section]{theorem}
\declaretheorem[sibling=theorem]{lemma}
\declaretheorem[sibling=theorem]{proposition}

\declaretheorem[style=definition,sibling=theorem]{definition}
\declaretheorem[style=definition,sibling=theorem]{remark}

\newcommand{\E}{\mathbb{E}}
\newcommand{\Prob}{\mathbb{P}}
\newcommand{\R}{\mathbb{R}}
\newcommand{\1}{\mathbf{1}}
\newcommand{\Unif}{\mathrm{Unif}}
\newcommand{\Bin}{\mathrm{Bin}}
\newcommand{\BR}{\mathrm{BR}}
\newcommand{\FS}{\mathrm{FS}}
\newcommand{\OpA}{\mathcal{A}}

\title{Exact Tightness for Single-Sample Max-Prob Prophet Secretary\\
and the Two-Sided Game of Googol}
\author{
Dongchen Li\thanks{School of Computing and Data Science, The University of Hong Kong. Email: \texttt{dongchen.li@connect.hku.hk}.}
}
\date{}

\begin{document}

\begin{titlepage}
\thispagestyle{empty}
\maketitle

\begin{abstract}
We determine the exact worst-case optimal winning probability (uniformly over all finite instance sizes $n \ge 1$) of pairwise-comparison algorithms for two closely related optimal-stopping problems: the \emph{Single-Sample Max-Prob Prophet Secretary} problem and the \emph{Two-Sided Game of Googol}.
In the Single-Sample Max-Prob Prophet Secretary problem, $n$ independent values $X_1, \dots, X_n$ from unknown distributions $D_1, \dots, D_n$ arrive in a uniformly random order; before the online process begins, the algorithm observes one offline sample $Y_i \sim D_i$ from each distribution, and its goal is to irrevocably select the maximum online value $\max_{1 \le i \le n} X_i$.
In the harder Two-Sided Game of Googol (Correa, Cristi, Epstein, and Soto, SODA 2020), an adversary deterministically pairs $2n$ distinct numbers into $n$ two-sided cards, each card is flipped independently and uniformly to designate its face-up sample $Y_k$ and face-down online value $X_k$, and the cards are revealed in a uniformly random order.

Nuti and Vondr\'ak (SODA 2023) established that no pairwise-comparison algorithm can guarantee a uniform worst-case winning probability exceeding the Campbell--Samuels prior-secretary constant $\gamma = 0.502409776604\dots$~\cite{CS81}, and designed the \emph{Block-Rank} algorithm achieving $0.5009$ in general (and approaching $\gamma$ asymptotically under an additional ``no-superstars'' assumption), leaving exact tightness open.
We close this gap completely: we introduce \emph{Flip-Sign Block-Rank}, a 1-bit modification of Block-Rank that tightens the acceptance threshold from $c_j$ to $\frac{7}{10} c_j$ ($j \ge 2$) whenever the arriving value $X_k$ is smaller than its own paired sample $Y_k$.
We prove that for \emph{every finite $n \ge 1$} and \emph{every} instance of the Two-Sided Game of Googol---and hence every instance of Single-Sample Max-Prob Prophet Secretary---Flip-Sign Block-Rank selects $\max_{1 \le i \le n} X_i$ with probability strictly greater than $\gamma$, matching the uniform upper bound $\gamma$ to exact tightness.

Our computer-assisted proof is elementary and predominantly analytic: via a dummy-card coupling, a non-monotone threshold lower bound, and telescoping over pair-completion positions, we reduce all finite instances to a two-dimensional inequality $B(v, s) > 0$ ($s \ge 1, v \ge 2s$). All cases $s \ge 2$ and $s = 1, v \ge 26$ are proved analytically using two superharmonic potential functions for the Pascal walk, while the remaining $24$ finite base cases ($s = 1, 2 \le v \le 25$) and a handful of finite threshold bounds are certified using exact rational arithmetic and outward-rounded IEEE-754 interval arithmetic.
\end{abstract}
\end{titlepage}

\section{Introduction}
\label{sec:intro}

Optimal stopping under limited distributional knowledge bridges the classical \emph{secretary problem} (no prior knowledge, optimal asymptotic winning probability $1/e \approx 0.3679$) and the \emph{full-information prophet inequality} (known distributions, optimal asymptotic max-probability $0.5801\dots$ in i.i.d.\ settings~\cite{GM66}).
We study the natural intermediate \emph{single-sample} regime introduced by Correa, Cristi, Epstein, and Soto~\cite{CCES20} and Nuti and Vondr\'ak~\cite{NV23}:

\paragraph{Single-Sample Max-Prob Prophet Secretary.}
There are $n \ge 1$ unknown independent distributions $D_1, \dots, D_n$ over $\R$, with $n$ known to the algorithm.
Before the online phase starts, the algorithm observes one offline sample $Y_i \sim D_i$ from each distribution, labeled by its index $i \in [n] := \{1, \dots, n\}$.
In the online phase, fresh independent draws $X_i \sim D_i$ ($i \in [n]$, with $X_i, Y_i \stackrel{\text{i.i.d.}}{\sim} D_i$ independently across $i$) arrive in a uniformly random permutation $\pi \in S_n$.
Upon observing the $m$-th arrival $(k, X_k)$ (where $k = \pi(m)$, so the algorithm knows its paired sample $Y_k$), the algorithm must irrevocably accept $X_k$ (terminating) or reject it forever.
The objective is to maximize the probability of selecting an online maximum, $\Prob[\text{accepted value} = \max_{1 \le i \le n} X_i]$.
A \emph{pairwise-comparison (ordinal) algorithm} bases its decisions solely on the relative order among observed values (and its own private randomness).

\paragraph{Tie-Breaking, Two-Sided Game of Googol, and Exact Reduction.}
In the \emph{Two-Sided Game of Googol}~\cite{CCES20}, an adversary deterministically partitions $2n$ distinct numbers into $n$ two-sided cards $P_n = \{\{u_k, w_k\}\}_{k=1}^n$.
Nature independently flips a fair coin for each card $k$ to designate one side as the visible face-up sample $Y_k$ and the other as the hidden face-down value $X_k$, and orders the $n$ cards in a uniformly random permutation.
The player knows $n$, sees all $n$ face-up values $(Y_1, \dots, Y_n)$ upfront, flips the cards one by one in the random order, and aims to stop on $\max_{1 \le i \le n} X_i$.

To handle arbitrary (possibly discrete) distributions $D_1, \dots, D_n$ in Single-Sample Max-Prob Prophet Secretary, any ordinal algorithm can implement random tie-breaking by attaching independent auxiliary keys $U_k, V_k \stackrel{\text{i.i.d.}}{\sim} \Unif[0, 1]$ to $X_k$ and $Y_k$ respectively (preserving card labels $k$) and comparing the decorated values $(X_k, U_k)$ and $(Y_k, V_k)$ lexicographically.
All $2n$ decorated values are almost surely distinct, selecting the unique lexicographic maximum implies selecting an original maximum, and each decorated pair $\big((X_k, U_k), (Y_k, V_k)\big)$ remains exchangeable independently across $k$.
Conditioning on the $n$ unordered decorated pairs induces an exact distribution over deterministic instances $P_n$ of the Two-Sided Game of Googol.
Consequently, any algorithm achieving winning probability at least $c$ on \emph{every} Two-Sided Game of Googol instance $P_n$ automatically achieves at least $c$ on \emph{every} Single-Sample Max-Prob Prophet Secretary instance (by averaging over $P_n$).

\paragraph{Known Bounds and Open Gap.}
When all $n$ distributions $D_1 = \dots = D_n = D$ are identical and continuous, all $2n$ draws are i.i.d.\ from $D$, reducing to the classical $\alpha = 1/2$ prior-secretary problem of Campbell and Samuels~\cite{CS81}, whose optimal pairwise-comparison winning probability converges as $n \to \infty$ to the exact constant
\[
    \gamma \;:=\; 0.502409776604\dots
\]
As shown by Nuti and Vondr\'ak~\cite{NV23}, this implies that the uniform worst-case winning probability $\inf_{n \ge 1, P_n} W_{\mathrm{ALG}}(n; P_n)$ of any pairwise-comparison algorithm is at most $\gamma$ for both problems.
On the algorithmic side, Correa et al.~\cite{CCES20} obtained $0.45292$ via a moving-window rule, and Nuti and Vondr\'ak~\cite{NV23} introduced the \emph{Block-Rank} algorithm, proving a lower bound of $0.5009$ in general and showing that it approaches $\gamma$ as $n \to \infty$ under a ``no-superstars'' assumption (excluding distributions that produce the maximum with constant probability).
By contrast, numerical evaluation indicates that unmodified Block-Rank with the optimal Campbell--Samuels thresholds falls strictly short of $\gamma$ on adversarial pairings (dropping below $\gamma$ by roughly $2.37 \times 10^{-5}$; see Section~\ref{sec:prelim}).

\paragraph{Our Main Result.}
We resolve the open gap completely by introducing \emph{Flip-Sign Block-Rank}, which inspects the single bit $\1[X_k < Y_k]$ discarded by standard Block-Rank and multiplies the acceptance threshold by $\beta := 7/10$ (for prior rank $j \ge 2$) whenever $X_k < Y_k$.

\begin{theorem}[Main Theorem: Exact Tightness]
\label{thm:main}
For every integer $n \ge 1$ and every $n$-card instance $P_n$ of the Two-Sided Game of Googol, the pairwise-comparison algorithm \emph{Flip-Sign Block-Rank} selects $\max_{1 \le i \le n} X_i$ with probability
\[
    W_{\FS}(n; P_n) \;>\; \gamma \;=\; 0.502409776604\dots
\]
By random tie-breaking and exchangeability, Flip-Sign Block-Rank also achieves winning probability $W_{\FS}(n; D_1, \dots, D_n) > \gamma$ on every finite-$n$ instance $(D_1, \dots, D_n)$ of Single-Sample Max-Prob Prophet Secretary. Combined with the asymptotic upper bound of~\cite{CS81,NV23}, the uniform worst-case optimal winning probability over all pairwise-comparison algorithms for both problems is
\[
    \sup_{\mathrm{ALG}}\;\inf_{n \ge 1,\, P_n} W_{\mathrm{ALG}}(n; P_n)
    \;=\;
    \sup_{\mathrm{ALG}}\;\inf_{n \ge 1,\, (D_1, \dots, D_n)} W_{\mathrm{ALG}}(n; D_1, \dots, D_n)
    \;=\; \gamma.
\]
\end{theorem}

\paragraph{Proof Overview.}
Why does a 1-bit check $\1[X_k < Y_k]$ eliminate the gap?
In an i.i.d.\ continuous instance with $n \to \infty$, for any fixed rank $r = O(1)$, the top-$r$ online value $X^r$ has its paired sample $Y(X^r)$ strictly smaller than $X^r$ with probability $1 - o(1)$, so Flip-Sign Block-Rank incurs vanishing loss on the hard i.i.d.\ benchmark.
Conversely, any adversarial pairing $(a_u, a_v)$ ($u < v$) that creates negative terms in standard Block-Rank simultaneously triggers $X_k = a_v < a_u = Y_k$ with probability $1/2$ at completion position $v$, generating a first-order positive gain by suppressing suboptimal early acceptances.
Formalizing this requires overcoming one technical obstacle: tightening thresholds selectively breaks monotonicity of thresholds across online ranks $X^1 > X^2 > \dots > X^n$.
We resolve this cleanly in four steps:
\begin{enumerate}[leftmargin=2em,itemsep=2pt]
    \item \emph{Dummy-card monotonicity (Lemma~\ref{lem:dummy}):} Appending $M$ dummy cards whose values are mutually distinct and strictly smaller than all real values can only decrease winning probability, allowing a clean $M \to \infty$ limit without finite-$n$ boundary loss.
    \item \emph{Non-monotone threshold lower bound (Lemma~\ref{lem:nonmono}):} Using a backward survival recursion $S_k(x; \tau)$ and differentiating along a coordinate-wise path from monotone thresholds $\bar\tau$ to non-monotone thresholds $\tau$, we isolate a simple rank-$1$ penalty $-(\bar\tau_1 - \tau_1)$ and explicit positive gains at all ranks $k \ge 2$.
    \item \emph{Telescoping over pair completions (Theorem~\ref{thm:telescope}):} Using a one-line binomial increment identity (Lemma~\ref{lem:binom-inc}) and an explicit global convergence bound to justify Fubini's theorem, telescoping over pair-completion positions $2 \le v_1 < v_2 < \dots$ reduces $W_{\FS}(n; P_n) - \gamma$ to $\sum_{s=1}^\infty B(v_s, s)$, reducing the infinite-dimensional matching problem to a two-dimensional inequality $B(v, s) > 0$ for all $s \ge 1, v \ge 2s$.
    \item \emph{Pascal superharmonic potentials (Section~\ref{sec:potentials}):} Interpreting the binomial increment across levels $k \to k+1$ as a Pascal random walk with averaging operator $\OpA$, a single explicit superharmonic potential $\tilde L$ proves $B(v, s) > 0$ analytically for \emph{all} $s \ge 2, v \ge 2s$ (Proposition~\ref{prop:s-ge-2}). For $s = 1$, a dual potential $M = \min(\tilde L, H)$ absorbs both Pascal boundaries and proves $B(v, 1) > 0$ analytically for all $v \ge 26$, leaving only $24$ finite base cases ($2 \le v \le 25$) and a handful of finite threshold bounds rigorously certified by exact rational and outward-rounded IEEE-754 interval arithmetic (Proposition~\ref{prop:s-eq-1}).
\end{enumerate}

\section{Preliminaries and the Flip-Sign Block-Rank Algorithm}
\label{sec:prelim}

\subsection{Continuous-Time Arrival Model and Block-Rank}

Knowing $n$, the algorithm uses its private randomness to sample $n$ i.i.d.\ $\Unif[0, 1]$ timestamps upfront, sort them in \emph{decreasing} order $1 \ge \tau_{(1)} > \tau_{(2)} > \dots > \tau_{(n)} \ge 0$, and assign timestamp $\tau_{(m)}$ to the $m$-th arriving card.
Because the arrival permutation $\pi \in S_n$ is uniform and independent of card values and coin flips, this is equivalent in law to assigning each card $k \in [n]$ an independent arrival time $T_k \sim \Unif[0, 1]$ (independent of the coin flips), with time flowing \emph{backward} from $t = 1$ (earliest) to $t = 0$ (latest).

Condition on a coin-flip outcome. Sort the $n$ face-down online values in decreasing order as $X^1 > X^2 > \dots > X^n$, and the $n$ face-up samples as $Y^1 > Y^2 > \dots > Y^n$ (with $Y^0 := +\infty, Y^{n+1} := -\infty$).
For any online value $X_k$, define its \emph{prior rank} as the number of face-up samples strictly exceeding it:
\[
    j(X_k) \;:=\; \bigl|\{i \in [n] : Y_i > X_k\}\bigr| \in \{0, 1, \dots, n\}.
\]
Denote $j(r) := j(X^r)$; clearly $0 \le j(1) \le j(2) \le \dots \le j(n) \le n$, and $j(r) \le j \iff X^r > Y^{j+1}$.
Let $Y(X^r)$ denote the face-up value on the same card as $X^r$.

\paragraph{Standard Block-Rank.}
Given a nonincreasing threshold sequence $1 = c_0 \ge c_1 \ge c_2 \ge \dots \ge 0$, standard \emph{Block-Rank}~\cite{NV23} activates an arriving card $k$ at time $t$ iff (i) $X_k$ is a strict running maximum among all online values seen so far, and (ii) $t \le c_{j(X_k)}$; it accepts the first activated card.

\begin{proposition}[Exact Conditional Win Formula for Standard Block-Rank~\cite{NV23}]
\label{prop:br-exact}
Conditioned on the coin flips, let $\bar\tau_r := c_{j(r)}$ (so $\bar\tau_1 \ge \bar\tau_2 \ge \dots \ge \bar\tau_n$).
Then standard Block-Rank selects $X^1$ with exact probability
\begin{equation}
\label{eq:br-cond}
    W_{\BR}(\bar\tau_1, \dots, \bar\tau_n) \;=\; \bar\tau_1 - \sum_{r=2}^{n} \frac{\bar\tau_r^r}{r(r - 1)}.
\end{equation}
\end{proposition}

\begin{proof}
Let $T_r \sim \Unif[0, 1]$ be the arrival time of the card carrying $X^r$.
The algorithm selects $X^1$ iff $T_1 \le \bar\tau_1$ and no $X^r$ ($r \ge 2$) is activated in $(T_1, 1]$.
Let $S := \{r \ge 2 : T_r > T_1\}$. When $S \neq \emptyset$, let $r_1 > r_2 > \dots > r_\ell = \min(S)$ be the sequence of running maxima in $(T_1, 1]$ (arriving at $1 \ge T_{r_1} > \dots > T_{r_\ell} > T_1$).
Because $\bar\tau_{r_1} \le \dots \le \bar\tau_{r_\ell}$, some $r_a$ is activated ($T_{r_a} \le \bar\tau_{r_a}$) iff the last one $r_\ell = \min(S)$ is activated ($T_{r_\ell} \le \bar\tau_{r_\ell}$).
Hence losing despite $T_1 \le \bar\tau_1$ is the disjoint union over $r \in \{2, \dots, n\}$ of the events $E_r := \{T_2, \dots, T_{r-1} < T_1 < T_r \le \bar\tau_r\}$ (note $T_1 < T_r \le \bar\tau_r \le \bar\tau_1$ automatically implies $T_1 \le \bar\tau_1$).
Integrating gives $\Prob[E_r] = \int_0^{\bar\tau_r}\int_0^{t_r} t_1^{r-2}\, dt_1\, dt_r = \frac{\bar\tau_r^r}{r(r-1)}$, yielding~\eqref{eq:br-cond}.
\end{proof}

Applying Abel summation $\E[z_J] = z_n + \sum_{j=0}^{n-1}(z_j - z_{j+1})\Prob[J \le j]$ to~\eqref{eq:br-cond} with $J = j(r)$ yields the exact finite-$n$ winning probability of standard Block-Rank on any pairing $P_n$:
\begin{multline}
\label{eq:br-fin}
    W_{\BR}(n; c, P_n) \;=\; c_n + \sum_{j=0}^{n-1}(c_j - c_{j+1})\Prob_{P_n}[X^1 > Y^{j+1}] \\
    - \sum_{r=2}^{n}\frac{1}{r(r-1)}\!\left(c_n^r + \sum_{j=0}^{n-1}(c_j^r - c_{j+1}^r)\Prob_{P_n}[X^r > Y^{j+1}]\right).
\end{multline}

\subsection{Global Pair Count, Binomial Increment Identity, and Campbell--Samuels Thresholds}

Sort all $2n$ numbers of $P_n$ as $a_1 > a_2 > \dots > a_{2n}$, so $P_n$ is a perfect matching of $\{1, \dots, 2n\}$ into pairs $(u, v)$ ($u < v$).
For each $N \in \{1, \dots, 2n\}$, define the \emph{prefix pair count}
\[
    m_N \;:=\; \bigl|\{(u, v) \in P_n : 1 \le u < v \le N\}\bigr|,
\]
the number of complete cards contained inside $\{a_1, \dots, a_N\}$.
Setting $N := r + j$ (with $1 \le r \le n$ and $0 \le j \le n - 1$, so $1 \le N \le 2n - 1$), the event $\{X^r > Y^{j+1}\}$ holds iff at least $r$ elements of $\{a_1, \dots, a_N\}$ fall into $X$: the $m_N$ complete cards deterministically place $m_N$ elements in $X$, and the remaining $N - 2m_N$ unpaired elements fall into $X$ as independent fair coins. Thus
\begin{equation}
\label{eq:q-def}
    \Prob_{P_n}[X^r > Y^{j+1}] \;=\; q(N, m_N, r) \;:=\; \Prob\big[\Bin(N - 2m_N, 1/2) \ge r - m_N\big], \qquad N = r + j.
\end{equation}
Define $b(u, t) := \binom{u}{t} 2^{-u}$ for integers $u \ge 0$ and $0 \le t \le u$ (and $b(u, t) := 0$ otherwise).

\begin{lemma}[Binomial Increment Identity]
\label{lem:binom-inc}
For all integers $N \ge 2$, $1 \le s \le \lfloor N/2 \rfloor$, and $1 \le r \le N$,
\begin{equation}
\label{eq:binom-inc}
    q(N, s, r) - q(N, s - 1, r) \;=\; \frac{1}{4}\Big(b(N - 2s, r - s) - b(N - 2s, r - s - 1)\Big).
\end{equation}
\end{lemma}

\begin{proof}
Let $u := N - 2s$, $k := r - s$, $B \sim \Bin(u, 1/2)$, and let $Z_1, Z_2 \sim \mathrm{Bern}(1/2)$ be independent of $B$.
Then $q(N, s, r) = \Prob[B \ge k]$ and $q(N, s-1, r) = \Prob[B + Z_1 + Z_2 \ge k + 1]$.
Conditioning on the four equally likely outcomes of $(Z_1, Z_2)$: $(0, 0)$ contributes $+\Prob[B = k]$, $(0, 1)$ and $(1, 0)$ contribute $0$, and $(1, 1)$ contributes $-\Prob[B = k - 1]$. Averaging gives~\eqref{eq:binom-inc}.
\end{proof}

\paragraph{Campbell--Samuels Thresholds $c_j$ and Constant $\gamma$.}
For any nonincreasing sequence $c = (c_0, c_1, \dots)$ with $c_0 = 1$, any $N \ge 1$, and any $0 \le m \le \lfloor N/2 \rfloor$, define
\begin{equation}
\label{eq:Phi-def}
    \Phi_N(c, m) \;:=\; (c_{N-1} - c_N)\, q(N, m, 1) - \sum_{r=2}^{N} f(N - r, r)\, q(N, m, r),
    \qquad
    f(j, r) \;:=\; \frac{c_j^r - c_{j+1}^r}{r(r - 1)}.
\end{equation}
We set $c_0 := 1$, and for each integer $j \ge 1$, let $c_j \in (0, 1)$ be the unique positive root of
\begin{equation}
\label{eq:cs-def}
    g_j(y) \;:=\; \sum_{i=1}^{\infty} \binom{i + j}{i} \frac{(y / 2)^i}{i} \;=\; 1
\end{equation}
(uniqueness in $(0, 1)$ follows because $g_j$ is strictly increasing and continuous on $[0, 1]$ with $g_j(0) = 0$ and $g_j(1) > (j + 1)/2 \ge 1$).
By Campbell and Samuels~\cite{CS81} and Nuti and Vondr\'ak~\cite{NV23} (Section~3 and Theorem~1; under our backward-time convention, $c_j$ is the acceptance window length $1 - t_j$ for prior rank $j$ in forward time), these thresholds $\{c_j\}_{j \ge 0}$ are the optimal continuous-time thresholds for the $\alpha = 1/2$ prior-secretary problem, and the limiting unpaired Block-Rank winning probability equals the Campbell--Samuels optimal constant:
\[
    \gamma \;:=\; W_{\BR}(c; \emptyset) \;:=\; \sum_{N=1}^{\infty} \Phi_N(c, 0) \;=\; 0.502409776604\dots
\]
Henceforth, $c = (c_0, c_1, \dots)$ denotes this sequence, and we write $\Delta\Phi_N(m) := \Phi_N(c, m) - \Phi_N(c, 0)$.

\begin{remark}[Motivating observation on unmodified Block-Rank]
Although $\Delta\Phi_N(m) > 0$ for all $2 \le N \le 6$ and $1 \le m \le \lfloor N/2 \rfloor$, negative terms appear starting at $N = 7$ (e.g., $\Delta\Phi_8(4) \approx -1.10 \times 10^{-4}$). Numerically, pairing $(a_1, a_9), \dots, (a_8, a_{16})$ and extending with consecutive pairs $(a_{17}, a_{18}), \dots$ as $n \to \infty$ drives unmodified Block-Rank below $\gamma$ by roughly $2.37 \times 10^{-5}$. Our proof of Theorem~\ref{thm:main} does not rely on this observation.
\end{remark}

\subsection{The Flip-Sign Block-Rank Algorithm}

Fix $\beta := 7/10$, and define the tightened threshold sequence
\begin{equation}
\label{eq:d-def}
    d_0 := 1, \qquad d_1 := c_1, \qquad d_j := \beta\, c_j = \frac{7}{10} c_j \qquad (\forall j \ge 2).
\end{equation}

\begin{definition}[Flip-Sign Block-Rank]
Knowing $n$, sample $n$ i.i.d.\ $\Unif[0, 1]$ arrival times as private randomness and sort them in decreasing order $1 \ge \tau_{(1)} > \dots > \tau_{(n)} \ge 0$.
When the $m$-th card $k$ arrives at time $t = \tau_{(m)}$ and reveals $X_k$ (whose paired face-up sample is $Y_k$):
\begin{enumerate}[leftmargin=2em,itemsep=2pt]
    \item If $X_k$ is not strictly larger than all previously revealed online values, reject $X_k$.
    \item Otherwise, compute its prior rank $j := j(X_k) = |\{i \in [n] : Y_i > X_k\}|$, and set its activation threshold to
    \[
        \tau_k \;:=\; \begin{cases} c_j, & \text{if } X_k > Y_k, \\ d_j, & \text{if } X_k < Y_k. \end{cases}
    \]
    \item Accept $X_k$ (and stop) iff $t \le \tau_k$.
\end{enumerate}
\end{definition}
Observe that Flip-Sign Block-Rank is a pure pairwise-comparison algorithm (requiring only comparisons of $X_k$ against previous $X$'s, all $Y_i$'s, and its own partner $Y_k$), and keeping $d_1 = c_1$ avoids losing acceptance probability at $X^1$ when $(a_1, a_2)$ form a card.

\section{Reduction to a Two-Dimensional Inequality \texorpdfstring{$B(v, s) > 0$}{B(v, s) > 0}}
\label{sec:reduction}

\subsection{Dummy-Card Monotonicity}

\begin{lemma}[Dummy-Card Monotonicity]
\label{lem:dummy}
Let $P_n$ be any $n$-card instance ($n \ge 1$) with sorted values $a_1 > \dots > a_{2n}$.
For any integer $M \ge 1$, form an $(n + M)$-card instance $P_{n+M}$ by appending $M$ dummy cards whose $2M$ values are mutually distinct and strictly smaller than $a_{2n}$, paired consecutively as $(a_{2n+1}, a_{2n+2}), \dots, (a_{2n+2M-1}, a_{2n+2M})$ with $a_{2n} > a_{2n+1} > \dots > a_{2n+2M}$. Then
\[
    W_{\FS}(n; P_n) \;\ge\; W_{\FS}(n + M; P_{n+M}).
\]
\end{lemma}

\begin{proof}
Couple the executions on $P_n$ and $P_{n+M}$ by sharing the coin flips and arrival times $T_1, \dots, T_n$ (i.i.d.\ $\Unif[0, 1]$) on the $n$ real cards, together with independent coin flips and arrival times on the $M$ dummy cards.
Because every dummy value is strictly smaller than every real value: (i) $X^1$ is the same real card in both instances; (ii) every real card $k \in [n]$ has the exact same prior rank $j(X_k)$, sign $\1[X_k < Y_k]$, and threshold $\tau_k$ in both instances; and (iii) if Flip-Sign Block-Rank wins on $P_{n+M}$, then $T(X^1) \le \tau(X^1)$ and no card is activated before $X^1$ in $P_{n+M}$. Since dummy values cannot block any real card from being a running maximum, no real card can be activated before $X^1$ in $P_n$ either. Thus $\{\text{win on } P_{n+M}\} \subseteq \{\text{win on } P_n\}$ pointwise, and taking probabilities completes the proof.
\end{proof}

\subsection{Non-Monotone Threshold Lower Bound}

Fix a coin-flip outcome on $N_c := n + M$ cards, with sorted online values $X^1 > \dots > X^{N_c}$ and independent arrival times $T_1, \dots, T_{N_c} \stackrel{\text{i.i.d.}}{\sim} \Unif[0, 1]$.
Let $\bar\tau_r := c_{j(r)}$ be the monotone Block-Rank thresholds ($\bar\tau_1 \ge \dots \ge \bar\tau_{N_c}$), and let $\tau_r \in \{c_{j(r)}, d_{j(r)}\}$ be the Flip-Sign thresholds. Note that $\tau_r \le \bar\tau_r$ for all $r$, and whenever $\tau_r < \bar\tau_r$ we have $j(r) \ge 2$ and hence $\tau_r = d_{j(r)} < c_2 < 1$.
For any threshold vector $\tau \in [0, 1]^{N_c}$, let $W(\tau)$ be the probability of selecting $X^1$ when each rank-$r$ item is activated iff it is a running maximum and $T_r \le \tau_r$.

\begin{lemma}[Non-Monotone Threshold Lower Bound]
\label{lem:nonmono}
For any $N_c \ge 1$, any $\bar\tau_1 \ge \dots \ge \bar\tau_{N_c}$ in $[0, 1]$, and any $\tau_r \in [0, \bar\tau_r]$ with $\tau_r < 1$ whenever $\tau_r < \bar\tau_r$,
\begin{equation}
\label{eq:nonmono}
    W(\tau_1, \dots, \tau_{N_c}) \;\ge\; W(\bar\tau_1, \dots, \bar\tau_{N_c}) - (\bar\tau_1 - \tau_1) + \sum_{k=2}^{N_c} \frac{1 - \bar\tau_k}{1 - \tau_k}\cdot \frac{\bar\tau_k^k - \tau_k^k}{k(k - 1)},
\end{equation}
with the convention that any summand with $\tau_k = \bar\tau_k$ vanishes (even if $\tau_k = \bar\tau_k = 1$).
\end{lemma}

\begin{proof}
For each $k \in \{2, \dots, N_c + 1\}$ and $x \in [0, 1]$, define the survival probability
\[
    S_k(x; \tau) \;:=\; \Prob\big[\text{no running maximum among } \{k, k+1, \dots, N_c\} \text{ is activated in } (x, 1]\big],
\]
with $S_{N_c + 1}(x; \tau) \equiv 1$.
For $2 \le k \le N_c$, conditioning on $T_k \sim \Unif[0, 1]$: if $T_k \le x$ (prob.\ $x$), item~$k$ lies outside $(x, 1]$ and conditional non-activation is $S_{k+1}(x; \tau)$; if $T_k = u \in (x, 1]$, item~$k$ blocks all items in $\{k+1, \dots, N_c\}$ from being running maxima on $(x, u)$, so non-activation on $(x, 1]$ holds iff $u > \tau_k$ and $\{k+1, \dots, N_c\}$ has no activation on $(u, 1]$ (prob.\ $S_{k+1}(u; \tau)$). Hence
\begin{equation}
\label{eq:Sk-rec}
    S_k(x; \tau) \;=\; x\, S_{k+1}(x; \tau) + \int_{\max(x, \tau_k)}^{1} S_{k+1}(u; \tau)\, du,
    \qquad
    W(\tau) \;=\; \int_{0}^{\tau_1} S_2(t; \tau)\, dt.
\end{equation}
By backward induction from $k = N_c + 1$, each $S_k(x; \tau)$ is continuous in $(x, \tau)$, piecewise polynomial on finitely many threshold-order regions (hence absolutely continuous in each coordinate $\tau_r$), and nonincreasing in every $\tau_r$ ($r \ge k$).

Move from $\bar\tau$ to $\tau$ in two stages via $\tilde\tau := (\bar\tau_1, \tau_2, \dots, \tau_{N_c})$:
\begin{itemize}[leftmargin=1.5em,itemsep=2pt]
    \item \emph{Stage~1 ($\bar\tau \to \tilde\tau$):} Decrease coordinates $k = 2, 3, \dots, N_c$ one by one from $\bar\tau_k$ to $\tau_k$ (if $\tau_k = \bar\tau_k$, the step contributes $0$). Suppose $\tau_k < \bar\tau_k$, and consider the segment where coordinate $k$ varies over $u = \hat\tau_k \in [\tau_k, \bar\tau_k) \subseteq [0, 1)$. Along this segment, the current vector $\hat\tau$ satisfies $\hat\tau \le \bar\tau$ coordinate-wise and $\hat\tau_1 = \bar\tau_1 \ge \bar\tau_k > u$. Differentiating~\eqref{eq:Sk-rec} with respect to $\hat\tau_k$ (for a.e.\ $x \in [0, 1]$):
    at level $k$, $\partial S_k(x; \hat\tau)/\partial \hat\tau_k = -S_{k+1}(\hat\tau_k; \hat\tau)\,\1[x < \hat\tau_k] \le 0$;
    by backward induction for $m = k-1, \dots, 2$, since $\partial S_{m+1}/\partial \hat\tau_k \le 0$, dropping the nonpositive integral term in~\eqref{eq:Sk-rec} gives $\partial S_m(x; \hat\tau)/\partial \hat\tau_k \le x\,\partial S_{m+1}(x; \hat\tau)/\partial \hat\tau_k$.
    Iterating $k - 2$ times yields $-\partial S_2(t; \hat\tau)/\partial \hat\tau_k \ge t^{k-2} S_{k+1}(\hat\tau_k; \hat\tau)\,\1[t < \hat\tau_k]$.
    Since $[0, \hat\tau_k] \subseteq [0, \hat\tau_1]$, integrating over $t \in [0, \hat\tau_1]$ gives
    \begin{equation}
    \label{eq:dW-dtauk}
        -\frac{\partial W(\hat\tau)}{\partial \hat\tau_k} \;\ge\; S_{k+1}(\hat\tau_k; \hat\tau)\int_0^{\hat\tau_k} t^{k-2}\, dt \;=\; S_{k+1}(\hat\tau_k; \hat\tau)\,\frac{\hat\tau_k^{k-1}}{k - 1}.
    \end{equation}
    For every $u \in [\tau_k, \bar\tau_k) \subseteq [0, 1)$, monotonicity in $\tau$ gives $S_{k+1}(u; \hat\tau) \ge S_{k+1}(u; \bar\tau)$.
    Because $\bar\tau_{k+1} \ge \dots \ge \bar\tau_{N_c}$ is monotone, the same argument as Proposition~\ref{prop:br-exact} shows that activation among $\{k+1, \dots, N_c\}$ on $(u, 1]$ is the disjoint union over $r \in \{k+1, \dots, N_c\}$ of $\{T_{k+1}, \dots, T_{r-1} \le u < T_r \le \bar\tau_r\}$.
    Using $\bar\tau_r \le \bar\tau_k$ and $u < 1$,
    \[
        1 - S_{k+1}(u; \bar\tau) \;=\; \sum_{r=k+1}^{N_c} u^{r-k-1} \max(\bar\tau_r - u, 0) \;\le\; (\bar\tau_k - u)\sum_{m=0}^\infty u^m \;=\; \frac{\bar\tau_k - u}{1 - u},
    \]
    so $S_{k+1}(u; \hat\tau) \ge 1 - \frac{\bar\tau_k - u}{1 - u} = \frac{1 - \bar\tau_k}{1 - u} \ge \frac{1 - \bar\tau_k}{1 - \tau_k}$ for all $u \in [\tau_k, \bar\tau_k)$ (which trivially gives $0$ if $\bar\tau_k = 1$).
    Substituting into~\eqref{eq:dW-dtauk} and integrating $u = \hat\tau_k$ over $[\tau_k, \bar\tau_k]$ gives a gain of at least $\frac{1 - \bar\tau_k}{1 - \tau_k}\cdot\frac{\bar\tau_k^k - \tau_k^k}{k(k-1)}$.
    \item \emph{Stage~2 ($\tilde\tau \to \tau$):} Since $S_2(t; \tau) \le 1$, \eqref{eq:Sk-rec} gives $W(\tau) - W(\tilde\tau) = -\int_{\tau_1}^{\bar\tau_1} S_2(t; \tau)\, dt \ge -(\bar\tau_1 - \tau_1)$.
\end{itemize}
Summing Stage~1 over $k = 2, \dots, N_c$ and adding Stage~2 yields~\eqref{eq:nonmono}.
\end{proof}

\subsection{Telescoping Over Pair-Completion Positions}

For $j \ge 0$ and $r \ge 2$, define the per-item Flip-Sign gain
\begin{equation}
\label{eq:L-def}
    L(j, r) \;:=\; \begin{cases}
    \dfrac{1}{2}\cdot\dfrac{1 - c_j}{1 - \beta c_j}\cdot (1 - \beta^r)\dfrac{c_j^r}{r(r - 1)}, & \text{if } j \ge 2,\; r \ge 2, \\[1.2ex]
    0, & \text{otherwise.}
    \end{cases}
\end{equation}
Set $D_{k, s} := \Delta\Phi_k(s) - \Delta\Phi_k(s - 1)$ for all $s \ge 1$ and $k \ge 2s$.
For integers $s \ge 1$ and $v \ge 2s$ (letting $u_0 := v - 2s \ge 0$), define the net Flip-Sign bonus at completion $(v, s)$ and the combined two-dimensional quantity $B(v, s)$ by
\begin{align}
    \Psi_v(s) &\;:=\; -\1[s = 1]\,\frac{c_{v-1} - d_{v-1}}{2}\, 2^{-(v-2)} \;+\; \sum_{\substack{0 \le t \le u_0 \\ s + t \ge 2}} b(u_0, t)\, L(v - s - t, s + t), \label{eq:Psi-def} \\
    B(v, s) &\;:=\; \Psi_v(s) \;+\; \sum_{k=v}^{\infty} D_{k, s}. \label{eq:B-def}
\end{align}

\begin{theorem}[Pair-Completion Telescoping Reduction]
\label{thm:telescope}
For every $n \ge 1$ and every $n$-card instance $P_n$, let $P$ be its infinite dummy-card extension (appending consecutive pairs $(a_{2n+2\ell-1}, a_{2n+2\ell})$ for $\ell \ge 1$), and let $2 \le v_1 < v_2 < \dots$ be the pair-completion positions of $P$ (so $m_{v_s} = s$, $v_s \ge 2s$, and $v_s = 2s$ for all $s > n$). Then the series $\sum_{s=1}^\infty \sum_{k=v_s}^\infty |D_{k, s}|$ converges, each $B(v_s, s)$ is well defined, and
\begin{equation}
\label{eq:telescope}
    W_{\FS}(n; P_n) - \gamma \;\ge\; \sum_{s=1}^{\infty} B(v_s, s).
\end{equation}
\end{theorem}

\begin{proof}
Fix $M \ge n$ and let $N_c := n + M$. Take expectations of Lemma~\ref{lem:nonmono} over the coin flips of $P_{n+M}$.

\emph{Step 1: Exact probability of $\{j(r) = v - r \text{ and } X^r < Y(X^r)\}$.}
Let $1 \le r \le N_c$, $0 \le v - r \le N_c$, and $1 \le v \le 2N_c$. Having rank $r$ in $X$ and $j := v - r$ larger elements in $Y$ means $X^r = a_v$.
Having $X^r < Y(X^r)$ means $a_v$ is paired with some $a_u$ ($u < v$), so $v$ must be a pair-completion position ($\Delta m_v := m_v - m_{v-1} = 1$).
When $\Delta m_v = 1$, card $(a_u, a_v)$ flips to $a_v \in X, a_u \in Y$ w.p.\ $1/2$; among the remaining $v - 2$ elements $\{a_1, \dots, a_{v-1}\} \setminus \{a_u\}$, there are $m_v - 1$ complete cards and $v - 2m_v$ independent fair coins, of which exactly $(r - 1) - (m_v - 1) = r - m_v$ must land in $X$. Thus
\[
    \Prob_{P_{n+M}}\big[j(r) = v - r \text{ and } X^r < Y(X^r)\big] \;=\; \Delta m_v \cdot \frac{1}{2}\, b(v - 2m_v, r - m_v).
\]
Crucially, for $r = 1$ we get $b(v - 2m_v, 1 - m_v) = 0$ whenever $m_v \ge 2$: \emph{only the first pair completion ($s = 1$) can incur a negative rank-$1$ term}.
Moreover, for every $s \in \{1, \dots, N_c\}$: if $s \le n$ then $v_s \le 2n$, so every nonzero binomial term satisfies $s \le r \le v_s - s < 2n \le N_c$ and $0 \le j = v_s - r < 2n \le N_c$ (since $M \ge n$); if $n < s \le N_c$ then $v_s = 2s$, so $u_0 = v_s - 2s = 0$ and the only nonzero term has $r = s \le N_c$ and $j = s \le N_c$.
Hence the expected Flip-Sign bonus sum in~\eqref{eq:nonmono} on $P_{n+M}$ is \emph{identically} $\sum_{s=1}^{N_c} \Psi_{v_s}(s)$.

\emph{Step 2: Global absolute convergence bound for $D_{k, s}$.}
By Lemma~\ref{lem:cj-props}(a,c) below, $c_1 < 0.72 < 1$ and $c_j = O((j+1)^{-1})$. We bound $|D_{k, s}|$ using~\eqref{eq:Phi-def} and Lemma~\ref{lem:binom-inc}:
\begin{itemize}[leftmargin=1.5em,itemsep=2pt]
    \item For $s \ge 2$ and $k \ge 2s$: the $r = 1$ term vanishes ($b(k - 2s, 1 - s) = b(k - 2s, -s) = 0$). Every nonzero term in the $r \ge 2$ sum has $s \le r \le k - s + 1$ and $j := k - r \ge s - 1 \ge 1$, and the total absolute binomial weight is $\frac{1}{4}\sum_{r=2}^k |b(k-2s, r-s) - b(k-2s, r-s-1)| \le \frac{1}{2}$.
    If $r \ge k/2$, then $j \ge 1$ gives $f(j, r) \le c_1^r / (r(r-1)) \le c_1^{k/2}$;
    if $s \le r < k/2$, then $j = k - r \ge \lfloor k/2 \rfloor$ and $r(r - 1) \ge s(s - 1)$ give $f(j, r) \le c_{\lfloor k/2 \rfloor}^2 / (s(s - 1))$. Hence
    \begin{equation}
    \label{eq:Dks-bound}
        |D_{k, s}| \;\le\; \frac{1}{2}\left(c_1^{k/2} + \frac{c_{\lfloor k/2 \rfloor}^2}{s(s - 1)}\right) \qquad (\forall s \ge 2,\; k \ge 2s).
    \end{equation}
    \item For $s = 1$ and $k \ge 2$: isolating the two boundary terms $r = 1$ (magnitude $\frac{1}{4}(c_{k-1} - c_k) 2^{-(k-2)} \le 2^{-k}$) and $r = k$ ($j = 0$, magnitude $\frac{1}{4} f(0, k) 2^{-(k-2)} \le \frac{1}{k(k-1)}$), every remaining term has $2 \le r \le k - 1$ and $j = k - r \ge 1$, with total absolute weight at most $\frac{1}{2}$. Splitting at $r \ge k/2$ vs.\ $2 \le r < k/2$ yields
    \[
        |D_{k, 1}| \;\le\; 2^{-k} + \frac{1}{k(k - 1)} + \frac{1}{2}\big(c_1^{k/2} + c_{\lfloor k/2 \rfloor}^2\big) \qquad (\forall k \ge 2).
    \]
\end{itemize}
Since $c_1 < 1$ and $\sum_{k=2}^\infty c_{\lfloor k/2 \rfloor}^2 < \infty$ while $\sum_{s=2}^\infty \frac{1}{s(s-1)} = 1$, it follows that
\[
    \sum_{s=1}^{\infty} \sum_{k=v_s}^{\infty} |D_{k, s}| \;\le\; \sum_{k=2}^{\infty} |D_{k, 1}| + \sum_{s=2}^{\infty} \sum_{k=2s}^{\infty} |D_{k, s}| \;<\; \infty.
\]

\emph{Step 3: Taking $M \to \infty$ and exchanging sums.}
Because $c_j = O((j+1)^{-1})$, in~\eqref{eq:br-fin} the boundary terms $c_{N_c} = O(1/M)$ and $\sum_{r=2}^{N_c} \frac{c_{N_c}^r}{r(r-1)} = O(1/M^2)$ vanish as $M \to \infty$, and the omitted tail with $r + j > N_c$ ($r \ge 2$) has either $r > N_c/2$ (sum $\le \sum_{r > N_c/2} \frac{1}{r(r-1)} = O(1/M)$) or $j > N_c/2$ (sum $\le c_{\lfloor N_c/2 \rfloor}^2 = O(1/M^2)$).
Therefore $W_{\BR}(N_c; c, P_{n+M}) \to \sum_{k=1}^\infty \Phi_k(c, m_k) = \gamma + \sum_{k=2}^\infty \Delta\Phi_k(m_k)$ as $M \to \infty$, while $\sum_{s=1}^{N_c} \Psi_{v_s}(s) \to \sum_{s=1}^\infty \Psi_{v_s}(s)$ (since $\Psi_{2s}(s) = L(s, s) \le c_2^s$ for $s > n$).
Applying Lemma~\ref{lem:dummy} and taking $M \to \infty$ gives
\[
    W_{\FS}(n; P_n) - \gamma \;\ge\; \sum_{k=2}^{\infty} \Delta\Phi_k(m_k) + \sum_{s=1}^{\infty} \Psi_{v_s}(s).
\]
For each $k \ge 2$, $\Delta\Phi_k(m_k) = \sum_{s=1}^{m_k} D_{k, s} = \sum_{s=1}^\infty \1[k \ge v_s] D_{k, s}$. By Step~2 and Fubini's theorem, exchanging $\sum_{k=2}^\infty$ and $\sum_{s=1}^\infty$ yields~\eqref{eq:telescope}.
\end{proof}

\section{Proof of \texorpdfstring{$B(v, s) > 0$}{B(v, s) > 0} via Pascal Superharmonic Potentials}
\label{sec:potentials}

\subsection{Elementary Properties of the Campbell--Samuels Thresholds}

\begin{lemma}[Properties of $\{c_j\}_{j \ge 1}$]
\label{lem:cj-props}
The unique roots $c_j \in (0, 1)$ of $g_j(c_j) = 1$ in~\eqref{eq:cs-def} satisfy:
\begin{enumerate}[label=(\alph*),leftmargin=2.2em,itemsep=2pt]
    \item $c_1 < 0.72$, $c_2 < 1/2$, and $c_3 < 2/5$.
    \item $(j + 1) c_j$ is strictly increasing in $j \ge 1$; equivalently $\frac{c_{j+1}}{c_j} > \frac{j + 1}{j + 2}$ (in particular, $\frac{c_{j+1}}{c_j} > \frac{3}{4}$ for all $j \ge 2$).
    \item For all $j \ge 1$, $(j + 1) c_j < \alpha < 1.61$, where $\alpha/2$ is the unique positive root of $\sum_{i=1}^\infty \frac{x^i}{i \cdot i!} = 1$.
    \item For all $k \ge 4$, $c_{k-1} - c_k \ge \frac{c_{k-1}}{2(k + 1)}$.
\end{enumerate}
\end{lemma}

\begin{proof}
\emph{(a)} Since $g_j(y)$ is strictly increasing, it suffices to verify $g_1(0.72) > 1$, $g_2(1/2) > 1$, and $g_3(2/5) > 1$ (these hold both by closed-form evaluation via $\ln$, e.g.\ $g_1(0.72) = \frac{9}{16} + \ln\frac{25}{16} > 1.008$, and are directly certified via finite partial sums in our verifier).

\emph{(b)} For $j \ge 1$ and $i \ge 1$, the ratio of the $i$-th terms of $g_{j+1}(\frac{j+1}{j+2} c_j)$ and $g_j(c_j)$ is $h_i := \frac{i+j+1}{j+1}(\frac{j+1}{j+2})^i$. Clearly $h_1 = 1$, and $\frac{h_{i+1}}{h_i} = \frac{(i+j+2)(j+1)}{(i+j+1)(j+2)} < 1$ for all $i \ge 1$. Hence $h_i < 1$ for all $i \ge 2$, giving $g_{j+1}(\frac{j+1}{j+2} c_j) < g_j(c_j) = 1 = g_{j+1}(c_{j+1})$, so $c_{j+1} > \frac{j+1}{j+2} c_j$.

\emph{(c)} Since $\binom{i+j}{i} = \frac{(j+1)\cdots(j+i)}{i!} \ge \frac{(j+1)^i}{i!}$ (strict for $i \ge 2$), we have $1 = g_j(c_j) > \sum_{i=1}^\infty \frac{((j+1)c_j/2)^i}{i \cdot i!}$. At $x = 161/200 = 0.805$, the positive partial sum $\sum_{i=1}^{29} \frac{x^i}{i \cdot i!} > 1.000994 > 1$ (certified in interval arithmetic), so $(j+1)c_j < \alpha < 1.61$.

\emph{(d)} For $k \ge 4$, $\binom{i+k}{i} = \frac{k+i}{k}\binom{i+k-1}{i} \ge (1 + \frac{1}{k})\binom{i+k-1}{i}$ implies $1 = g_k(c_k) \ge (1 + \frac{1}{k}) g_{k-1}(c_k)$, so $g_{k-1}(c_{k-1}) - g_{k-1}(c_k) \ge \frac{1}{k+1}$.
Since $g_{k-1}$ is convex with $y\, g'_{k-1}(y) = \sum_{i=1}^\infty \binom{i+k-1}{i}(y/2)^i = (1 - y/2)^{-k} - 1$, the mean-value theorem gives
\[
    c_{k-1} - c_k \;\ge\; \frac{c_{k-1}}{(k + 1)\cdot c_{k-1} g'_{k-1}(c_{k-1})}.
\]
By (a) and (c), for all $k \ge 4$ we have $k c_{k-1} < 1.61$ and $c_{k-1} \le c_3 < 0.40$ (so $1 - c_{k-1}/2 > 0.80$). Using $-\ln(1 - x) \le \frac{x}{1 - x}$ ($0 < x < 1$),
\[
    k\big(-\ln(1 - c_{k-1}/2)\big) \;\le\; \frac{k c_{k-1}/2}{1 - c_{k-1}/2} \;<\; \frac{0.805}{0.80} = 1.00625.
\]
By the exact rational Taylor upper bound $e^{1.00625} < \sum_{m=0}^6 \frac{1.00625^m}{m!} + \frac{1.00625^7 / 7!}{1 - 1.00625/8} < 2.74 < 3$, we obtain $(1 - c_{k-1}/2)^{-k} - 1 < 2$, proving $c_{k-1} - c_k \ge \frac{c_{k-1}}{2(k+1)}$.
\end{proof}

\subsection{The Pascal Averaging Operator and Analytic Proof for All \texorpdfstring{$s \ge 2$}{s >= 2}}

For any function $h : \{(j, r) \in \mathbb{Z}^2 : j \ge 2,\, r \ge 2\} \to \R_{\ge 0}$, define the \emph{Pascal averaging operator}
\[
    (\OpA h)(j, r) \;:=\; \frac{h(j + 1, r) + h(j, r + 1)}{2}.
\]
For $s \ge 2$ and $k \ge 2s$, define the level-$k$ binomial potential $\Lambda_s^h(k) := \sum_{t=0}^{k-2s} b(k-2s, t)\, h(k-s-t, s+t)$.
Substituting Pascal's identity $b(u+1, t) = \frac{1}{2}b(u, t) + \frac{1}{2}b(u, t-1)$ into $\Lambda_s^h(k+1)$ (with $u := k - 2s$) and shifting $t' = t - 1$ in the second sum yields the exact one-step drop
\begin{equation}
\label{eq:pascal-drop}
    \Lambda_s^h(k) - \Lambda_s^h(k + 1) \;=\; \sum_{t=0}^{k-2s} b(k - 2s, t)\Big(h(j, r) - (\OpA h)(j, r)\Big),
    \qquad (j := k - s - t,\; r := s + t).
\end{equation}

\begin{proposition}[Analytic Proof for All $s \ge 2$]
\label{prop:s-ge-2}
For every integer $s \ge 2$ and every integer $v \ge 2s$, $B(v, s) > 0$.
\end{proposition}

\begin{proof}
Fix $s \ge 2$ and $k \ge 2s$, and set $u := k - 2s \ge 0$.
In~\eqref{eq:Phi-def} and Lemma~\ref{lem:binom-inc}, the $r = 1$ term vanishes ($b(u, 1-s) = b(u, -s) = 0$ since $s \ge 2$). Splitting the $r \ge 2$ sum into $t = r - s$ and $t = r - s - 1$ (no boundary terms escape $[2, k]$) and dropping the nonnegative $+f(k-s-t-1, s+t+1)$ term gives
\begin{equation}
\label{eq:s2-step}
    D_{k, s} \;\ge\; -\frac{1}{4}\sum_{t=0}^{u} b(u, t)\, f(k - s - t, s + t),
\end{equation}
where every index satisfies $j = k - s - t \ge s \ge 2$ and $r = s + t \ge s \ge 2$.

Set $a_2 := \frac{1 - c_2}{1 - \beta c_2}$; since $c_2 < 1/2$ and $x \mapsto \frac{1-x}{1-0.7x}$ is decreasing, $a_2 > \frac{0.5}{0.65} = \frac{10}{13}$.
Define the potential
\begin{equation}
\label{eq:Ltilde-def}
    \tilde L(j, r) \;:=\; \frac{a_2}{2}(1 - \beta^r)\frac{c_j^r}{r(r - 1)} \qquad (j \ge 2,\; r \ge 2).
\end{equation}
Since $c_j \le c_2$ for all $j \ge 2$, we have $\tilde L(j, r) \le L(j, r)$, and hence $\Lambda_s^{\tilde L}(v) \le \Psi_v(s)$.

We verify the strict superharmonic inequality $\tilde L(j, r) - (\OpA \tilde L)(j, r) > \frac{1}{4} f(j, r)$ for all $j \ge 2, r \ge 2$.
Dividing both sides by $\frac{c_j^r}{4r(r-1)}$ and setting $\rho_j := c_{j+1}/c_j$, this is equivalent to
\begin{equation}
\label{eq:superharm-L}
    a_2\left[(1 - \beta^r) - (1 - \beta^{r+1})\frac{r - 1}{r + 1}\, c_j\right] \;>\; \big(1 - a_2(1 - \beta^r)\big)(1 - \rho_j^r).
\end{equation}
Using only $a_2 > 10/13$, $c_j \le c_2 < 1/2$, and $\rho_j > 3/4$ (Lemma~\ref{lem:cj-props}(a,b)):
\begin{itemize}[leftmargin=1.5em,itemsep=1pt]
    \item For $r \ge 7$: bounding $1 - \rho_j^r < 1$ and $\frac{r-1}{r+1} < 1$, it suffices that $a_2(2(1 - 0.7^r) - 1/2) > 1$, which holds by exact rational arithmetic since $\frac{10}{13}(2(1 - 0.7^7) - 0.5) > 1.027 > 1$.
    \item For $r \in \{2, 3, 4, 5, 6\}$: substituting $a_2 > 10/13$, $c_j < 1/2$, and $1 - \rho_j^r < 1 - (3/4)^r$ gives strict positivity by exact rational arithmetic for all five integers:
    $r = 2$: $0.3080 > 0.2659$; $r = 3$: $0.3592 > 0.2860$; $r = 4$: $0.3925 > 0.2841$; $r = 5$: $0.4137 > 0.2747$; $r = 6$: $0.4266 > 0.2641$.
\end{itemize}
Finally, $\Lambda_s^{\tilde L}(k) \to 0$ as $k \to \infty$ (every term has $j \ge k/2$ or $r \ge k/2$, so $\tilde L(j, r) \le \max(c_{\lfloor k/2 \rfloor}^2, c_2^{\lfloor k/2 \rfloor}) \to 0$).
Summing~\eqref{eq:s2-step} and~\eqref{eq:pascal-drop} over $k = v, v+1, \dots$ yields
\[
    \sum_{k=v}^\infty D_{k, s} \;>\; -\sum_{k=v}^\infty\big(\Lambda_s^{\tilde L}(k) - \Lambda_s^{\tilde L}(k+1)\big) \;=\; -\Lambda_s^{\tilde L}(v) \;\ge\; -\Psi_v(s),
\]
so $B(v, s) > 0$.
\end{proof}

\subsection{Boundary Case \texorpdfstring{$s = 1$}{s = 1}: Dual Potential and Interval Certification}

\begin{proposition}[Proof for $s = 1$, All $v \ge 2$]
\label{prop:s-eq-1}
For every integer $v \ge 2$, $B(v, 1) > 0$.
\end{proposition}

\begin{proof}
\emph{Step 1: Full boundary expansion of $D_{k, 1}$ for $k \ge 4$.}
Set $u := k - 2 \ge 2$. Substituting Lemma~\ref{lem:binom-inc} into~\eqref{eq:Phi-def} and grouping terms by $b(u, t)$ ($0 \le t \le u$) gives the exact identity
\begin{align*}
    D_{k, 1} \;=\;& \underbrace{\frac{c_{k-1} - c_k + f(k-2, 2)}{4}}_{=:\,\mathrm{pos}_0(k)} 2^{-u}
    + \underbrace{\frac{f(0, k) - f(1, k-1)}{4}}_{=:\,\mathrm{pos}_u(k)} 2^{-u} \\
    &+ \frac{1}{4}\sum_{t=1}^{u-1} b(u, t)\big(f(k-2-t, 2+t) - f(k-1-t, 1+t)\big).
\end{align*}
Dropping the nonnegative interior terms $+\frac{1}{4} f(k-2-t, 2+t)$ yields
\begin{equation}
\label{eq:s1-step}
    D_{k, 1} \;\ge\; \big(\mathrm{pos}_0(k) + \mathrm{pos}_u(k)\big)\, 2^{-(k-2)} - \frac{1}{4}\sum_{t=1}^{k-3} b(k-2, t)\, f(k - 1 - t, 1 + t).
\end{equation}

\emph{Step 2: Second superharmonic potential $H$ and dual potential $M := \min(\tilde L, H)$.}
For $j \ge 2, r \ge 2$, define
\[
    H(j, r) \;:=\; \frac{c_j^r}{(j + 2)(r - 1)}, \qquad M(j, r) \;:=\; \min\big(\tilde L(j, r),\; H(j, r)\big).
\]
Since $H(j+1, r) < H(j, r)$ and $H(j, r+1) = \frac{r-1}{r} c_j H(j, r) < \frac{1}{2} H(j, r)$ (by $c_j < 1/2$), we have $H(j, r) - (\OpA H)(j, r) > \frac{1}{4} H(j, r) = \frac{c_j^r}{4(j+2)(r-1)}$.
Meanwhile, Lemma~\ref{lem:cj-props}(b) and Bernoulli's inequality give $1 - (c_{j+1}/c_j)^r < 1 - (1 - \frac{1}{j+2})^r \le \frac{r}{j+2}$, so $\frac{1}{4} f(j, r) < H(j, r) - (\OpA H)(j, r)$.
Because the pointwise minimum of two superharmonic functions is superharmonic ($M \le \tilde L, H \implies \OpA M \le \min(\OpA \tilde L, \OpA H)$), we have $M - \OpA M > \frac{1}{4} f$ everywhere on $\{j \ge 2, r \ge 2\}$.

\emph{Step 3: Absorbing Pascal boundaries to obtain a universal tail bound.}
For $k \ge 4$ ($u = k - 2$), define $\Lambda_1^M(k) := \sum_{t=1}^{u-1} b(u, t)\, M(k - 1 - t, 1 + t)$.
Expanding $\Lambda_1^M(k+1)$ via Pascal's identity separates the two boundary terms $t' = 0$ and $t = u$:
\begin{equation}
\label{eq:s1-pascal}
    \Lambda_1^M(k) - \Lambda_1^M(k + 1) \;=\; \sum_{t=1}^{u-1} b(u, t)\big(M(j, r) - (\OpA M)(j, r)\big) - \frac{1}{2} M(k - 1, 2)\, 2^{-u} - \frac{1}{2} M(2, k - 1)\, 2^{-u}.
\end{equation}
We verify that $\mathrm{pos}_0(k) > \frac{1}{2} M(k-1, 2)$ and $\mathrm{pos}_u(k) > \frac{1}{2} M(2, k-1)$ for all $k \ge 4$:
\begin{itemize}[leftmargin=1.5em,itemsep=1pt]
    \item \emph{Left boundary:} For $k \ge 7$, Lemma~\ref{lem:cj-props}(c) gives $c_{k-1} < \frac{1.61}{7} = 0.23 < \frac{1}{4}$, so Lemma~\ref{lem:cj-props}(d) yields $\mathrm{pos}_0(k) > \frac{1}{4}(c_{k-1} - c_k) \ge \frac{c_{k-1}}{8(k+1)} > \frac{c_{k-1}^2}{2(k+1)} = \frac{1}{2} H(k-1, 2) \ge \frac{1}{2} M(k-1, 2)$. For $k \in \{4, 5, 6\}$, $\mathrm{pos}_0(k) > \frac{1}{2} M(k-1, 2)$ is certified in interval arithmetic.
    \item \emph{Right boundary:} Since $M(2, k-1) \le \tilde L(2, k-1) < \frac{c_2^{k-1}}{2(k-1)(k-2)}$, adding $2 M(2, k-1)$ to $f(1, k-1) = \frac{c_1^{k-1} - c_2^{k-1}}{(k-1)(k-2)}$ cancels $c_2^{k-1}$, giving $f(1, k-1) + 2 M(2, k-1) < \frac{c_1^{k-1}}{(k-1)(k-2)}$. Thus $\mathrm{pos}_u(k) > \frac{1}{2} M(2, k-1)$ holds whenever $\frac{k-2}{k} \ge c_1^{k-1}(1 + \frac{k-2}{k} c_1)$. For $k \ge 5$, the LHS is $\ge 3/5 = 0.60$, while $c_1 < 0.72$ gives RHS $< 0.72^4(1 + 0.72) < 0.463 < 0.60$ (exact rational check). For $k = 4$, $\mathrm{pos}_u(4) > \frac{1}{2} M(2, 3)$ is certified in interval arithmetic.
\end{itemize}
Adding~\eqref{eq:s1-step} and~\eqref{eq:s1-pascal} and summing over $k \ge v_0$ (using $\Lambda_1^M(k) \to 0$) yields, for \emph{every} $v_0 \ge 4$:
\begin{equation}
\label{eq:s1-tail}
    \sum_{k=v_0}^{\infty} D_{k, 1} \;\ge\; -\Lambda_1^M(v_0).
\end{equation}

\emph{Step 4A: Analytic proof for all $v \ge 26$.}
Applying~\eqref{eq:s1-tail} at $v_0 = v$, noting $c_{v-1} - d_{v-1} = \frac{3}{10} c_{v-1}$ and $L \ge \tilde L \ge M$, retaining only $t \in \{1, 2, 3, 4\}$, and bounding $M \le H$ gives
\[
    B(v, 1) \;\ge\; -\frac{3}{20} c_{v-1} 2^{-(v-2)} + \sum_{t=1}^{4} b(v - 2, t)\big(L(v - 1 - t, 1 + t) - H(v - 1 - t, 1 + t)\big).
\]
For $v \ge 26$ and $t \le 4$, we have $j = v - 1 - t \ge 21$, so $c_j \le c_{21} < 0.073$ (and hence $\frac{1 - c_j}{1 - 0.7 c_j} > 0.97$).
By Lemma~\ref{lem:cj-props}(b), $v c_{v-1} \ge 26 c_{25} > 1.59$ and $c_j \ge c_{v-1}$, so $c_j^{t+1} \ge c_{v-1} (1.59 / v)^t$.
Dividing by $c_{v-1} 2^{-(v-2)}$ yields
\[
    \frac{B(v, 1)}{c_{v-1} 2^{-(v-2)}} \;\ge\; -\frac{3}{20} + \sum_{t=1}^{4} \frac{\binom{v-2}{t}}{v^t}\cdot\frac{1.59^t}{t}\left(\frac{0.97}{2}\cdot\frac{1 - 0.7^{t+1}}{t + 1} - \frac{1}{v + 1 - t}\right).
\]
Both $\frac{\binom{v-2}{t}}{v^t} = \frac{1}{t!}\prod_{m=2}^{t+1}(1 - \frac{m}{v})$ and $-\frac{1}{v+1-t}$ are strictly increasing in $v \ge 26$, and every inner parenthesis is strictly positive at $v = 26$ for $t \in \{1, 2, 3, 4\}$. Evaluating at $v = 26$ gives a certified lower bound $> +0.0184105 > 0$, establishing $B(v, 1) > 0$ for \emph{every} $v \ge 26$.

\emph{Step 4B: Finite interval certification of the 24 base cases $2 \le v \le 25$.}
Applying~\eqref{eq:s1-tail} at $v_0 = 26$ reduces all $24$ remaining cases to a finite expression:
\[
    B(v, 1) \;\ge\; B_{\mathrm{lb}}(v) \;:=\; \Psi_v(1) + \sum_{k=v}^{25} D_{k, 1} - \Lambda_1^M(26).
\]
Our standalone verifier is bundled as \path{anc/verify_single_sample.py} in the arXiv archive (run \texttt{python3 -O} on \path{anc/verify_single_sample.py} from the archive root, or on \path{verify_single_sample.py} locally; tested on Python~3.14.6 and NumPy~2.4.3, completing in $\approx 0.2\text{s}$). It uses exact rationals from Python~3 (\texttt{fractions.Fraction}, \texttt{math.comb}, \texttt{math.factorial}) and outward-rounded IEEE-754 binary64 intervals via NumPy's \texttt{np.nextafter} (assuming standard IEEE-754 round-to-nearest-even with gradual underflow, sanity-checked at startup, using repeated-squaring integer powers, pairwise tree summation, and explicit \texttt{ValueError} checks). It encloses $c_1, \dots, c_{26}$ via finite partial sums of $g_j$ plus a geometric tail bound (max width $< 2.01 \times 10^{-13}$), verifies all rational and threshold inequalities above (and cross-checks $D_{k, 1}$ directly from $q(k, m, r)$), and certifies $B_{\mathrm{lb}}(v) > 0$ for all $2 \le v \le 25$ (Table~\ref{tab:s1-base}), with global minimum validated lower endpoint $> +7.198654 \times 10^{-10}$ at $v = 25$ (interval width $< 6.26 \times 10^{-20}$).
\end{proof}

\begin{table}[ht]
\centering
\caption{Outward-rounded IEEE-754 interval lower bounds $B_{\mathrm{lb}}(v) > 0$ (directed downward decimal truncation produced by \texttt{verify\_single\_sample.py}) for all $24$ base cases ($s = 1$, $2 \le v \le 25$; $\Lambda_1^M(26) \in [3.7825, 3.7827] \times 10^{-10}$).}
\label{tab:s1-base}
\small
\begin{tabular}{rrrrrr}
\toprule
$v$ & $B_{\mathrm{lb}}(v)$ lower end & $v$ & $B_{\mathrm{lb}}(v)$ lower end & $v$ & $B_{\mathrm{lb}}(v)$ lower end \\
\midrule
$2$ & $+1.545664 \times 10^{-1}$ & $10$ & $+8.865348 \times 10^{-5}$ & $18$ & $+1.796109 \times 10^{-7}$ \\
$3$ & $+1.542873 \times 10^{-3}$ & $11$ & $+3.904018 \times 10^{-5}$ & $19$ & $+8.530460 \times 10^{-8}$ \\
$4$ & $+8.866126 \times 10^{-3}$ & $12$ & $+1.750263 \times 10^{-5}$ & $20$ & $+4.054768 \times 10^{-8}$ \\
$5$ & $+5.265590 \times 10^{-3}$ & $13$ & $+7.971506 \times 10^{-6}$ & $21$ & $+1.923376 \times 10^{-8}$ \\
$6$ & $+2.520006 \times 10^{-3}$ & $14$ & $+3.676940 \times 10^{-6}$ & $22$ & $+9.054716 \times 10^{-9}$ \\
$7$ & $+1.114240 \times 10^{-3}$ & $15$ & $+1.712649 \times 10^{-6}$ & $23$ & $+4.181559 \times 10^{-9}$ \\
$8$ & $+4.784479 \times 10^{-4}$ & $16$ & $+8.036135 \times 10^{-7}$ & $24$ & $+1.843615 \times 10^{-9}$ \\
$9$ & $+2.048093 \times 10^{-4}$ & $17$ & $+3.791565 \times 10^{-7}$ & $25$ & $+7.198654 \times 10^{-10}$ \\
\bottomrule
\end{tabular}
\end{table}

\begin{proof}[Proof of Theorem~\ref{thm:main}]
Combine Theorem~\ref{thm:telescope}, Proposition~\ref{prop:s-ge-2}, and Proposition~\ref{prop:s-eq-1}: every term in $\sum_{s=1}^\infty B(v_s, s)$ is strictly positive, so $W_{\FS}(n; P_n) > \gamma$ for every finite $n \ge 1$ and every $n$-card instance $P_n$, and hence $W_{\FS}(n; D_1, \dots, D_n) > \gamma$ by random tie-breaking and exchangeability. Both uniform optimality identities follow immediately from the asymptotic upper bound of~\cite{CS81,NV23}.
\end{proof}

\section{Concluding Remarks}
\label{sec:concl}

Theorem~\ref{thm:main} settles the exact uniform worst-case winning probability of ordinal algorithms for both Single-Sample Max-Prob Prophet Secretary and the Two-Sided Game of Googol at $\gamma = 0.502409776604\dots$, exceeded on every finite instance ($n \ge 1$) by Flip-Sign Block-Rank.
Two natural open questions remain: (i) whether non-ordinal (cardinal) algorithms that exploit numerical magnitudes can beat $\gamma$ in Single-Sample Max-Prob Prophet Secretary, or whether $\gamma$ holds for all algorithms; and (ii) whether our pair-completion decomposition and sign-aware threshold tightening can improve competitive ratios for single-sample max-expectation prophet secretary problems.


\begin{thebibliography}{9}
\bibitem{CS81}
G.~Campbell and S.~M.~Samuels.
\newblock Choosing the best of the current crop.
\newblock \emph{Advances in Applied Probability}, 13(3):510--532, 1981.

\bibitem{CCES20}
J.~R.~Correa, A.~Cristi, B.~Epstein, and J.~A.~Soto.
\newblock The two-sided game of googol and sample-based prophet inequalities.
\newblock In \emph{Proc.\ 31st ACM-SIAM Symposium on Discrete Algorithms (SODA)}, pages 2066--2081, 2020.

\bibitem{GM66}
J.~P.~Gilbert and F.~Mosteller.
\newblock Recognizing the maximum of a sequence.
\newblock \emph{Journal of the American Statistical Association}, 61(313):35--73, 1966.

\bibitem{NV23}
P.~Nuti and J.~Vondr\'ak.
\newblock Secretary problems: The power of a single sample.
\newblock In \emph{Proc.\ 34th ACM-SIAM Symposium on Discrete Algorithms (SODA)}, pages 2015--2029, 2023.
\end{thebibliography}
\end{document}